\documentclass[final,5p,times,twocolumn]{elsarticle}

\usepackage{graphics}

\usepackage{amssymb}

\journal{Nuclear Instruments and Methods A}

\begin{document}

\begin{frontmatter}



\title{High level of $^3$He polarization of 81\% Maintained in an on-beam $^3$He spin filter using
SEOP}


\author{E. Babcock}
\author{S. Mattauch}
\author{A. Ioffe}

\address{J$\ddot{u}$lich Centre for Neutron Science, Instiut f$\ddot{u}$r Festk$\ddot{o}$rperforshung, Forschungszentrum J$\ddot{u}$lich GmbH, Lichtenberg Str. 1, Garching, Germany}

\begin{abstract}
Maintaining high levels of $^3$He polarization over long periods
of time is important to many areas of fundamental and particle
beam physics. Long measurement times are often required in such
experiments and the data quality is a function of the $^3$He
polarization. This is the case for neutron scattering where the
$^3$He can be used to analyze the spin of a scattered neutron beam
and relatively small fluxes of polarized neutrons leads to
experiment times longer than several days. Consequently the
J$\rm\ddot{u}$lich Centre for Neutron Science (JCNS) is developing
spin-exchange optical pumping (SEOP) systems capable of polarizing
the $^3$He gas in place on a typical neutron instrument. Using a
polarizer device we constructed a high level of $^3$He
polarization of 81 \% $\pm$2 \%was maintained with good time
stability. Such levels of polarization maintained over time will
be able to reduce the measurement times for such experiments and
eliminate time dependent data corrections.
\end{abstract}

\begin{keyword}
polarized 3He \sep neutron spin filter \sep spin-exchange optical
pumping

\end{keyword}

\end{frontmatter}


\section{Introduction}
Polarized $^3$He has been shown to be advantageous when used as an
analyzer for polarized neutron scattering, an area of physics
research that is growing quickly given its relevance to studies of
magnetism and also soft matter. These sorts of experiments are
normally conducted in large scale facilities for neutron
scattering research where even at the highest flux neutron sources
these experiments can require many days of data collection because
the measurements are commonly counting statistics limited. Factors
such as small sample size and low solid angle coverage of the
scattered neutron beam can lead to long experiments. Additional
losses in neutron flux caused by the addition of neutron
polarizers and analyzers, and also because one now must normally
measure four spin states, 2 $\times$ 2, for the two states of
incoming neutron spin, and the two states of scattered neutron
spin, for each measurement further lengthens required measurement
times.

Much work has been done to optimize the use of polarized $^3$He
for use as neutron spin filters (NSF).  Routinely experiments
using $^3$He NSF are conducted at many neutron sources worldwide.
These experiments include polarized neutron diffraction, polarized
neutron reflectometry, polarized small angle neutron scattering
and fundamental particle physics \cite{ Chupp2007, Andersen2009,
Petoukhov2006, Chen2009, Gentile2005b}. Further, many neutron
sources worldwide have programs developing the methods and
acquiring the technology for $^3$He NSF locally for use at their
facilities. At the JCNS we have chosen to work on in-situ
polarization of the $^3$He gas using the spin exchange optical
pumping method (SEOP) \cite{ Walker97}.  $^3$He polarizations of
up to 80\% in $^3$He NSF cells have been verified with neutron
measurements \cite{Andersen2009, Chen2009, Parnell2009}. In these
experiments the $^3$He was polarized off-line, and thus undergoes
T$_1$ nuclear magnetic decay of the $^3$He polarization which
typically has a time constant of 100-300 hours. Here we present a
measurement of 80\% $^3$He polarization but maintained in steady
state for 24 hours on a neutron instrument by in-situ optical
pumping.
\section{In-situ SEOP polarizer}
\begin{figure}
\includegraphics[scale=0.4]{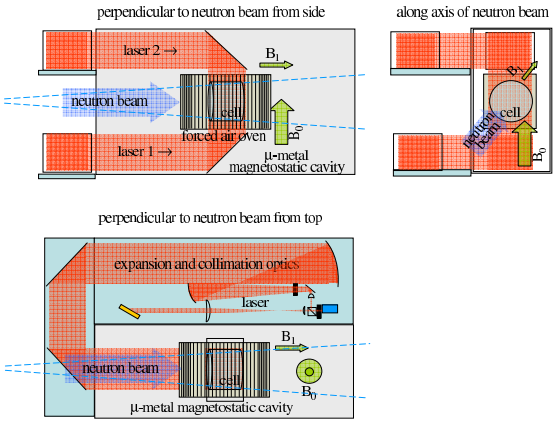}
\caption{Diagram of the in-situ SEOP polarizer showing the
configuration of the magnetic cavity, lasers and optics, RF coil
for adiabatic fast passage, oven and cell. \label{diagram}}
\end{figure}

A typical in-situ SEOP polarizer must contain several typical
elements. First a magnetic environment with a uniformity on the
order of 10$^{-4}$ cm$^{-1}$ in $\Delta B/B$ is required to
maintain long $^3$He T$_1$ lifetimes. Second, a circularly
polarized high power laser source and collimation optics are
required for the optical pumping of the dense alkali-metal vapor
in the cell. Next a heating source/oven is needed to obtain the
necessary alkali-metal number densities which requires
temperatures ranging from 170 $^{\circ}C$ to 250 $^{\circ}C$. And
optionally a transverse RF field for performing adiabatic fast
passage spin reversal of the $^3$He can be installed so that the
system can be used also be used as a neutron flipper. Several such
SEOP based polarizer have been constructed for NSF \cite{
Chupp2007,Jones2006, Ino2009, Lee2010}A block diagram of our
device is shown in Figure \ref{diagram}. Additionally, diagnostics
such as a free induction decay system to monitor $^3$He
polarization independent of neutrons and pump light absorption can
be installed to aid in system monitoring and optimization. The
total length of this device is about 1 meter including an
enclosure for laser radiation protection. This device is in many
ways similar to the one presented in \cite{ Boag2008,
Babcock2008}.

Saturation polarization in a SEOP cell is determined from the
balance of the alkali-metal $^3$He spin exchange rate and the
total $^3$He relaxation rate for a certain set of conditions where
the spin-exchange rate to the $^3$He is a function of the number
density of the alkali-metal vapor. Typically the maximum
polarization in SEOP cells is limited by a $^3$He relaxation term,
called the X-factor, that scales also with alkali-metal number
density \cite{Babcock2006}. While the origins of this relaxation
are still not understood, they seem to be a fixed property of a
given cell. Typical cells used for NSF have an X-factor of 0.2 to
0.3 times the spin-exchange rate. This leads to maximum $^3$He
polarization in NSF of 70\% to 80\% when one includes the other
sources of $^3$He relaxation such as relaxation from collisions
with the cell walls and dipole-dipole self-relaxation. These
latter two effects are added together and cited as the measured
room temperature lifetime, or T$_1$, for the cell. The cell we
used, named J1 here and in prior publications, was chosen for this
test because it has a low X-factor value and a long T$_1$ of 660
hours. This cell was produced in the Forschungszentrum-J$\rm
\ddot{u}$lich GmbH glass workshop and prepared and filled with
rubidium and $^3$He in a collaboration with ISIS. It has been
characterized in previous neutron tests, where it was polarized
off-line, and showed a high level of $^3$He polarization
\cite{Parnell2009}, but was never polarized in-situ. Consequently
those measurements involved extrapolation of the $^3$He
polarization to the maximum equilibrium value.

In addition narrow band optical pumping sources, with spectral
linewidth comparable to the pressure broadened absorption
linewidth of the Rb vapor in the SEOP cell, are necessary to
obtain the highest levels of alkali-metal polarization
\cite{Lancor2010}. Narrow band optical pumping sources must
therefore be used to obtain maximum $^3$He polarization because it
is also proportional to the average alkali-metal polarization in
equilibrium. Consequently the lasers that we choose to use are
frequency narrowed using an external cavity \cite{ Babcock2005a}
which can meet the requirements for polarizing Rb vapor to near
unity. The laser must then be expanded to the cover the full
dimensions of the cell and collimated to provide the most optimal
performance \cite{ Chann2002b}.
\section{Measurement}
The device was installed on the TREFF magnetic reflectometer
\cite{TREFF}. It was placed directly after the neutron sample
position and electromagnet such that the cell was approximately 90
cm from the sample position. The magnet was run at a nominal 0.01
T field during this test. A single neutron counter was placed
behind the system and used to monitor the neutron transmission of
the unpolarized incident beam through the $^3$He cell as a
function of time.

The relative neutron transmission, $i.e.$ the ratio of
transmission, $T$, of an unpolarized neutron beam through a
polarized $^3$He cell to the transmission through the unpolarized
cell, $T_0$, is
\begin{equation}
{T\over {T_0}}={\rm cosh([He]}l\sigma P_{\rm He}\lambda).
\end{equation}
Here $\rm [He]$ is the $^3$He number density, $\sigma$ the spin
dependent $^3$He neutron absorption cross section, $\lambda$ the
neutron wavelength, $l$ the cell length, and $P_{\rm He}$ the
$^3$He polarization. Thus by knowing the $^3$He number
density-length-cross section product, and the neutron wavelength,
unpolarized neutron transmission is a strait forward method to
obtain absolute $^3$He polarization. The neutron wavelength of
TREFF has been previously calibrated and is 4.7 \AA $\pm$ 0.1 \AA.

The $^3$He number density-length-cross section product of a $^3$He
cell can be determined from the unpolarized neutron transmission
of an unpolarized $^3$He cell as a function of neutron wavelength.
The unpolarized neutron transmission of an unpolarized $^3$He
cell, $T_0$ is
\begin{equation}
T_0(\lambda)={T_{\rm E}e^{{\rm [He]}l\sigma\lambda}}.
\end{equation}
Thus by measuring the transmission as a function of neutron
wavelength one obtains a two parameter fit where the empty cell
transmission is $T_{\rm E}$ and the value of the exponential is
the number density-length-cross section product. In this way the
$^3$He number density-length-cross section product of the J1 cell
was measured with a small angle neutron scattering spectrometer
KWS1 \cite{KWS1} which has been installed at the FRM II reactor in
Garching. KWS1 uses a 10\% $\Delta\lambda/\lambda$ neutron
velocity selector to change the neutron wavelength. From this
measurement the number density length product of the J1 cell was
found to be 5.18 bar$\cdot$cm $\pm$ 0.08 bar$\cdot$cm where the
pressure in bar is referenced to 25 $^{\circ}$C to convert to a
number density.

\begin{figure}
\includegraphics[scale=0.7]{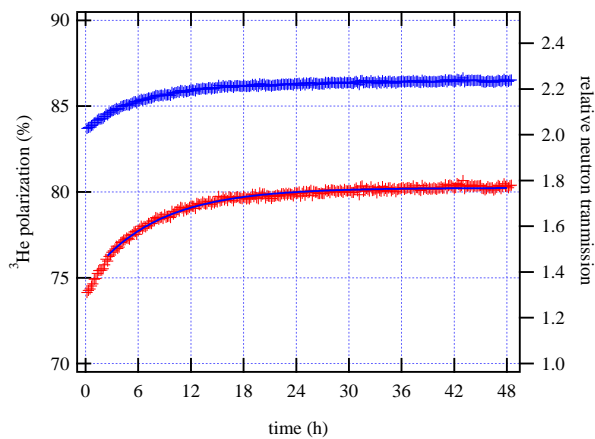}
\caption{The $^3$He polarization versus time for the J1 cell, red
(left axis) and the neutron transmission relative the unpolarized
cell transmission, blue (right axis). A fit to the data gives a
spin-exchange time constant for the build up of $^3$He
polarization of 7 hours. \label{polarization}}
\end{figure}

After optimization of the optical pumping parameters of our
system, during which time the cell had already achieved over 74\%
$P_{\rm He}$, it was allowed to run continuously for two days
without any additional adjustments. During this time the neutron
transmission was monitored with the neutron detector. Figure
\ref{polarization} shows the measured relative neutron
transmission which has been normalized to the unpolarized cell
transmission versus time. The unpolarized $^3$He cell transmission
was obtained directly at the end of the measurement by
depolarizing the cell with an intense RF field at the $^3$He
Larmour frequency. Polarization destruction was verified by a
nuclear magnetic resonance free induction decay system installed
to monitor the relative $^3$He polarization independent of the
neutron data.

From this data one can see the final transmission of the polarized
cell was a factor of 2.23 times the unpolarized transmission
corresponding to an equilibrium $^3$He polarization of 81\% $\pm$
2 \%. The polarizer was highly stable for the period of the 48
hour measurement. A fit to the data gives a 7 hour spin-exchange
time constant which is a typical value for the SEOP method and the
absolute polarization was over 80\% for the last 24 hours of the
measurement.

The in-situ $^3$He relaxation time of the J1 cell for this
experiment was measured before the final polarization measurement
by turning off the lasers and the cell heating while monitoring
neutron transmission. This data was again normalized to the
unpolarized neutron transmission, and the fit gave T$_1$=351 hours
$\pm$ 4 hours. The reduction of the cell T$_1$ from 660 hours to
351 hours is caused by the field homogeneity in our $\rm\mu$-metal
cavity being lower when installed on TREFF than what one obtains
in laboratory conditions. This adds some additional $^3$He
relaxation du to the field gradients. However, despite this
reduction in T$_1$, it was sufficiently long compared to the
spin-exchange rate not to affect the maximum achievable
polarization within the accuracy of our measurement.
\section{Conclusion}
In conclusion, we have measured and maintained a new high-level of
$^3$He polarization for a continuously polarized NSF on a neutron
instrument. Being able to maintain this level of polarization will
serve to improve the quality of polarized neutron instrumentation
using polarized $^3$He.
\section{Acknowledgments}
We greatly acknowledge the contributions of P. Bush for assistance
with the neutron measurements on KWS1, S. Boag and S. Parnell for
help filling the J1 cell, S. Staringer for aiding with the
mechanical design of the polarizer, the technical staff at the
JCNS, and the workshop in the Institut f$\rm\ddot{u}$r
Festk$\rm\ddot{o}$rperforschung (IFF) and the glass workshop at
the Zentralabteilung Technologie (ZAT) both in the
Forschungszentrum J$\rm\ddot{u}$lich GmbH. E. Babcock would also
like to acknowledge the $^3$He polarization group at the Instutut
Laue Langevin (ILL) in Grenoble (France) where he was a post-doc
from 2005 to 2008 working on similar techiques and devices.


\bibliographystyle{elsarticle-num}
\bibliography{BabcockBib80nim}







\end{document}